\documentclass[aps,prl,superscriptaddress,amsmath,amssymb,floatfix,twocolumn]{revtex4-1}
\usepackage{times}
\usepackage{graphicx}
\usepackage{subfigure}
\usepackage{color}
\usepackage{dsfont}
\usepackage{relsize}
\usepackage{epstopdf}

\begin{document}

\title{
Orbital-selective superconductivity
in the nematic phase of FeSe }

\author{Haoyu Hu}
\email{hh25@rice.edu}
\affiliation{Department of Physics \& Astronomy, Center for Quantum Materials, Rice University, Houston, Texas 77005,USA}

\author{Rong Yu}
\email{rong.yu@ruc.edu.cn}
\affiliation{Department of Physics and Beijing Key Laboratory of Opto-electronic Functional Materials and Micro-nano Devices, Renmin University of China, Beijing 100872, China}

\author{Emilian M. Nica}
\email{enica@asu.edu}
\affiliation{Department of Physics, Box 871504, Arizona State University, Tempe, Arizona  85287-1504}

\author{Jian-Xin Zhu}
\email{jxzhu@lanl.gov}
\affiliation{Theoretical Division and Center for Integrated Nanotechnologies, Los Alamos National Laboratory, Los Alamos, New Mexico 87545, USA}

\author{Qimiao Si}
\email{qmsi@rice.edu}
\affiliation{Department of Physics \& Astronomy, Center for Quantum Materials, Rice University, Houston, Texas 77005,USA}

\begin{abstract}
 The interplay between electronic orders and superconductivity is central to the physics of
 unconventional
superconductors, and is particularly pronounced in the iron-based superconductors.
Motivated by recent experiments on FeSe, we study the superconducting pairing
in its nematic phase 
in a multiorbital model with
 frustrated spin-exchange interactions. 
Electron correlations
in the presence of nematic order
give rise to an enhanced orbital selectivity in the
superconducting pairing amplitudes.
This orbital-selective pairing produces a large gap anisotropy
on the Fermi surface.
Our results naturally explain the
striking experimental observations,
and shed light on the unconventional superconductivity of correlated electron systems
in general.
\end{abstract}
\maketitle


{\it Introduction.~}
High temperature superconductivity in the iron-based superconductors (FeSCs) is
a major frontier of condensed matter physics \cite{Hos18,NatRevMat:2016,Dai2015}.
New 
phenomena and insights continue to arise  in this area, giving hope for deep understandings of
the ingredients that are central to the mechanism of superconductivity.
One such ingredient is the orbital-selective Mott physics
\cite{NatRevMat:2016,MYi.2016}.
It has been
 advanced for multiorbital models of the FeSCs \cite{Yu2011,Yu2013,deMedici2014},
in which the lattice symmetry dictates
 the presence of 
interorbital kinetic hybridizations,
and has been
observed by angle-resolved photoemission spectroscopy (ARPES)
\cite{MYi.2013,MYi.2015,YJPu.2016,MYi.2016}.
The orbital-selective Mott physics connects well with the 
bad-metal normal state \cite{Si2008,Haule08}, 
as implicated by the room-temperature electrical resistivity reaching the Mott-Ioffe-Regel limit
and the 
Drude weight 
having a large correlation-induced reduction
\cite{Basov.2009}.
Another closely related 
ingredient is orbital-selective superconducting pairing (OSSP),
which was initially advanced for the purpose of understanding the gap anisotropy of iron-pnictide
superconductors \cite{Yu_PRB2014}.

Among the FeSCs, the bulk FeSe system is of particular interest. 
It 
is the structural basis 
of the single-layer FeSe
on an SrTiO$_3$ substrate, which holds the record for 
the  superconducting transition temperature $T_c$
in the FeSCs \cite{Xue.2012,SLHe2013,Shen.2014,YWang2015}.
It has a nematic ground state, which reduces the $C_4$ rotational symmetry of a tetragonal lattice to $C_2$ and in turn
lifts the degeneracy between the $d_{xz}$ and $d_{yz}$ orbitals.

More generally,  FeSe provides a setting to study the interplay between the orbital selectivity and  electronic orders.
Indeed, recent scanning tunneling microscopy (STM) measurements 
in the nematic phase of FeSe
have 
uncovered a surprisingly large
difference between the quasiparticle weights of
the $d_{xz}$ and $d_{yz}$ orbitals, suggesting the proximity to the orbital-selective Mott phase \cite{Davis_2018}.
Moreover, they suggest a strongly orbital-selective
superconducting state,
as reflected in an unusually large anisotropy of
the superconducting gap \cite{stm}:  The ratio of the maximum to the minimum of the gap, $\Delta_{\text{max}}/\Delta_{\text{min}}$, 
is at least 
about $4$.
Recently, several of us have suggested a microscopic picture for the orbital-selective Mott physics in the nematic 
but 
normal (i.e., non-superconducting) state
\cite{Rong2018}. 
Within a slave-spin approach,
electron correlations in the presence of
 nematic order
 are found to yield a large
difference in  the quasiparticle weights of the $d_{xz}$ and $d_{yz}$ orbitals while the
associated band-splittings as seen in ARPES are relatively small \cite{arpes,XJZhou}.

In this Rapid Communication, we study the pairing structure in the nematic phase of FeSe 
using this theoretical picture.
We show that the orbital selectivity in the normal state leads
to an orbital-selective pairing,
which in turn produces a large gap anisotropy that is consistent with the STM results.
Our work not only provides a natural understanding of the experimental observations, but also sheds light
on the interplay between the orbital-selective pairing/Mott physics and electronic orders, all of which appear to be
important ingredients 
for the unconventional superconductivity in FeSCs and beyond.

{\it Model and method.~}
As a starting point, we consider the five-orbital Hubbard model for FeSe. The Hamiltonian reads as $H=H_{t}+H_{\rm{int}}$. 
Here, $H_{t}=\sum_{ij,\alpha\beta}t_{ij}^{\alpha\beta} c_{i,\alpha,\sigma}^\dag c_{j,\beta,\sigma}$,
where $c_{i,\alpha,\sigma}^\dag$ creates an electron in orbital $\alpha(\in{xz,yz,x^2-y^2,xy,z^2})$, spin $\sigma$
and at site $i$ of an Fe-square lattice.
The tight-binding parameters are
obtained by fitting the {\em ab initio} density functional theory (DFT) bandstructure of FeSe,
and 
$H_{\rm{int}}$ describes the on-site interactions,
which include the intra- and inter-orbital Coulomb repulsions
and the Hund's coupling [see Supplemental Material (SM) \cite{suppl}].
We use the $U(1)$ slave-spin method~\cite{Yu_PRB2012,Yu_PRB2017}
to study the correlation effects of this model.
In this representation,
the electron creation operator is expressed
as $c_{i,\alpha,\sigma}^\dag =S^{+}_{i,\alpha,\sigma} f_{i,\alpha,\sigma}^\dag $,
where $S^+_{i,\alpha,\sigma}$
is the ladder operator of a quantum $S=1/2$ slave spin
and $f_{i,\alpha,\sigma}^\dag $ is the creation operator of a fermionic spinon.
The effective strength of the correlation effect in orbital $\alpha$ is characterized by the quasiparticle spectral weight
$Z_\alpha \sim |\langle S_\alpha^\dag\rangle|^2$ (here we have dropped the site and spin indices).
$Z_\alpha>0$ describes the spectral weight for the coherent itinerant electrons,
 while $Z_\alpha=0$ refers to a Mott localization of the corresponding orbital.
We obtain $Z_\alpha$ for each orbital in the 
nematic 
normal
(i.e., non-superconducting) 
 state via solving the slave-spin
saddle-point equations detailed in Refs.~\onlinecite{Yu_PRB2012,Yu_PRB2017}.
Calculations in Ref.~\cite{Rong2018} for nematic normal state
 yield a strongly orbital-dependent
spectral weight of the order
$Z_{xz}:Z_{yz}:Z_{xy}=1:4:0.5$,
which is consistent with the values extracted from the STM measurements \cite{Davis_2018,stm,Kreisel2017}.
We will adopt this ratio for our calculation.
An important advantage of the $U(1)$ slave-spin approach in comparison with, for instance,
 the $Z_2$ counterpart \cite{deMedici05,Ruegg2010,Nandkishore2012},
 is that the slave-spin operators can carry all the charge degrees of freedom
 and the $f$ fermions are left with carrying all the spin degrees of freedom.
Consequently, in the bad-metal regime,
we can get a low-energy effective model
by integrating out the incoherent part of the electron spectrum (via the
quantum fluctuation of the slave spins) \cite{NatRevMat:2016,Ding_arXiv2014,Dai_PNAS2009}.
The resulting effective model
can be written in terms of the
$f$-fermion operators as follows,
\begin{eqnarray}\label{HamtJ}
H_{eff}&=&\sum_{ij,\alpha\beta}(\sqrt{Z_{\alpha}Z_{\beta}}t_{ij}^{\alpha\beta}-\lambda_{\alpha}\delta_{\alpha\beta})
f_{i,\alpha,\sigma}^\dag f_{j,\beta,\sigma}\nonumber\\
&&-\sum_{ij,\alpha\beta} J_{ij}^{\alpha\beta}f_{j,\beta,\downarrow}^\dag f_{i,\alpha,\uparrow}^\dag f_{i,\alpha,\downarrow}f_{j,\beta,\uparrow}.
\end{eqnarray}
It takes the form of a multiorbital $t$-$J$ model with the spin-exchange couplings
$J_{ij}^{\alpha\beta}$
coming from the integrating-out procedure.
The slave-spin calculations for the renormalization factors, $Z_{\alpha}$ for orbital $\alpha$,
are similar to those for the normal nematic state of FeSe
as described in Ref.\,\onlinecite{Rong2018},
with a bare Coulomb interaction being about 3.5 eV.
The intraorbital components $J_{1}^{\alpha}$ and $J_{2}^{\alpha}$, for  the nearest neighbor $\langle ij \rangle$ and 
next nearest-neighbor $\langle\langle ij \rangle\rangle$, 
will be used.

To study the superconductivity,
we define the pairing amplitude of the $f$ fermions to be
$\widetilde{\Delta}^{\alpha\beta}_{\textbf{e}}=
\frac{1}{2N}\sum_{i}
\langle f_{i,\alpha,\uparrow}f_{i+\textbf{e},\beta,\downarrow}-
f_{i,\alpha,\downarrow}f_{i+\textbf{e},\beta,\uparrow}\rangle$
refers to a unit vector connecting nearest and next nearest neighboring sites.
We
treat
the four-fermion $J$ terms through a Hubbard-Stratonovich decoupling, and self-consistently solve
the pairing amplitudes $\widetilde{\Delta}^{\alpha\beta}_{\textbf{e}}$
in
 the resulting effective model.
The pairing amplitude of the physical
electrons $\Delta_{\textbf{e}}^{\alpha\beta}=\frac{1}{2N}\sum_{i}
\langle c_{i,\alpha,\uparrow} c_{i+\textbf{e},\beta,\downarrow}-
c_{i,\alpha,\downarrow} c_{i+\textbf{e},\beta,\uparrow}\rangle$ is
\begin{eqnarray}\label{pairing}
{\Delta}_{\textbf{e}}^{\alpha\beta} &=&
\sqrt{Z_{\alpha}Z_{\beta}} \widetilde{\Delta}_{\textbf{e}}^{\alpha\beta}.
\end{eqnarray}

\begin{figure}[t!]
\centering\includegraphics[
width=50mm
]{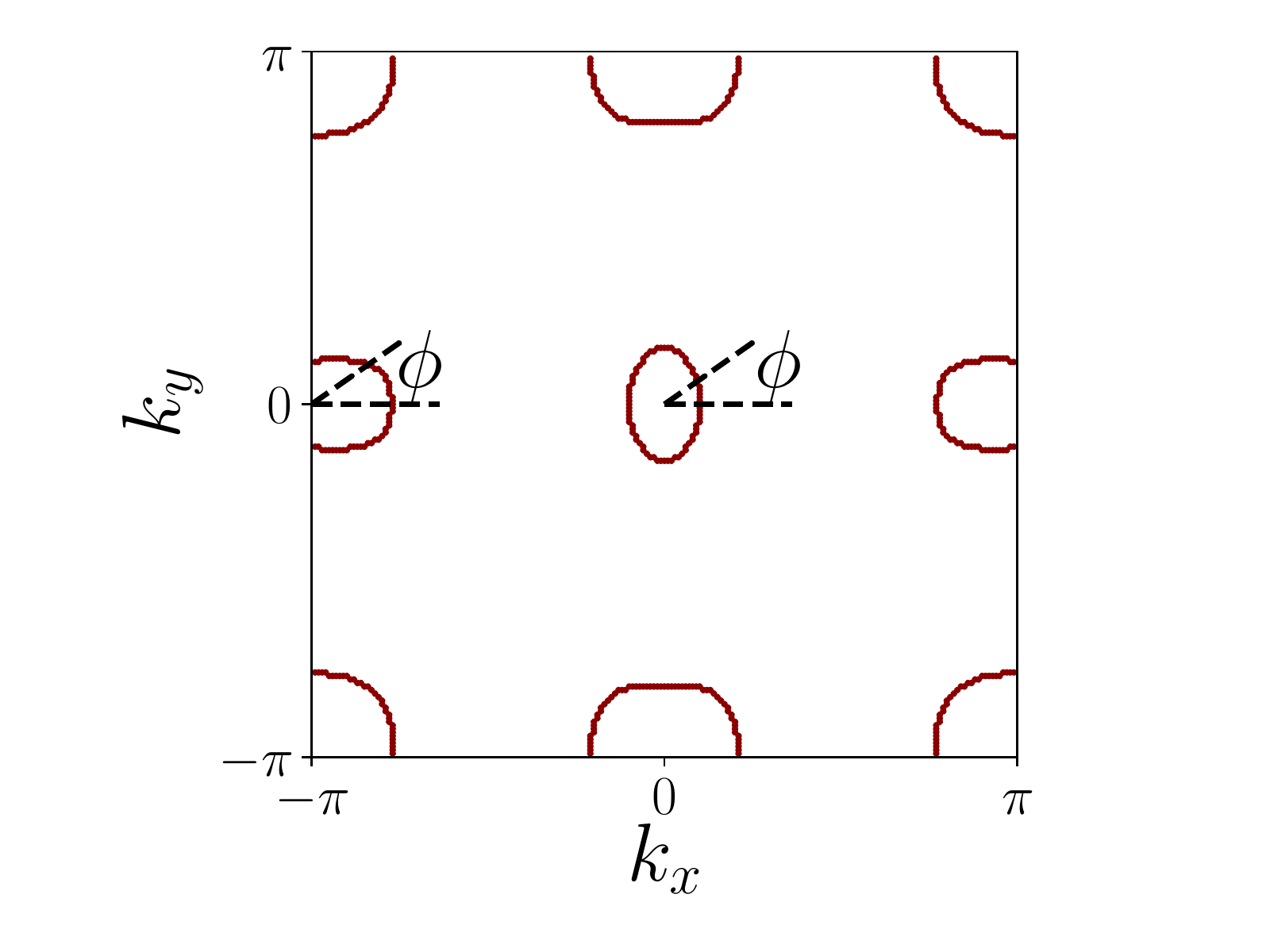}
\caption{Calculated Fermi surface in the 
nematic 
normal
phase of FeSe
with
$\eta=0.07$ and 
$\lambda = -0.03$.
}
\label{fig:1}
\end{figure}

{\it Nematic order.~} 
In the nematic phase, the breaking of $C_4$
symmetry
induces additional anisotropies
to both the kinetic energy and exchange interactions.
To take this effect into account 
in a simple way, we introduce an anisotropy parameter $\eta$
in the nearest-neighbor hopping parameters and the exchange couplings 
of the $d_{xz/yz}$ orbitals
as follows,
\begin{eqnarray}\label{nematicity}
t_{x/y}
=t(1\pm \eta); ~~~~~~
J_{x/y}
=J(1\pm \eta)^2.
\end{eqnarray}
For example, the nearest-neighbor hopping terms 
of the $d_{xz/yz}$ orbitals 
contains  the following 
in the nematic phase,
\begin{eqnarray}\label{oo2}
 &&\eta \left[ t_1(c_{xz,i}^\dag c_{xz,i+e_x}-c_{yz,i}^\dag c_{yz,i+e_y})\right.\nonumber\\
& +& \left.  t_2(-c_{xz,i}^\dag c_{xz,i+e_y}+c_{yz,i}^\dag c_{yz,i+e_x})\right].\nonumber
 \end{eqnarray}
The latter corresponds to a combination of
the $s$- and $d$-wave
bond nematic orders \cite{bond_nematic_waston},
\begin{eqnarray}\label{oo3}
&&\eta \left[\frac{t_1-t_2}{2}(\cos(k_x)+\cos(k_y))(n_{xz,k}-n_{yz,k})\right.\nonumber\\
&& \left. + \frac{t_1+t_2}{2}(\cos(k_x)-\cos(k_y))(n_{xz,k}+n_{yz,k})\right] . \nonumber
\end{eqnarray}

{\it Fermi surface in the nematic phase.~}
We use the
notation of the 1-Fe Brillouin zone (BZ).
In Fig.~\ref{fig:1},
we show the Fermi surface 
in the nematic phase 
for $\eta=0.07$.
An atomic spin-orbit coupling (SOC),
of the form $\lambda \textbf{S} \cdot \textbf{L}$,
is included in the calculation for 
Fig.\,\ref{fig:1}.
The superconductivity considered here is mainly driven by the magnetic interactions. 
Because the SOC is much smaller than the magnetic bandwidth,
its effect on the pairing will be neglected.
With increasing $\eta$, the inner hole pocket near the $\Gamma$ point
quickly 
disappears;
this evolution is shown in
Fig.S1 of the SM \cite{suppl}.
The (outer) hole pocket near the $\Gamma$ point
is elongated along the $k_y$ direction. 
The electron pocket near the $M_x$ [$(\pi,0)$] point is also elongated, along the $k_x$ direction.
 The electron pocket is dominated by the $d_{yz}$ and $d_{xy}$ orbitals,
 whereas the hole pocket mainly comprises the $d_{xz}$ and $d_{yz}$ orbitals (Fig. S2 \cite{suppl}).
The hole pocket near the $(\pi,\pi)$ point,
which appears in our model 
as a result of the known artifact of the DFT calculations 
 \cite{dft2008,dft2010,dft_jpsj},
does not come into play in
our main result.

\begin{center}
\begin{table}[t!]
\centering
  \begin{tabular}{ | l | l | l | l | }
    \hline
      pairing channel &  $D_{4h}$  &  $D_{2h}$   &   pairing channel in real space \\ \hline
    $s_{x^2+y^2}\ \tau_0$   &   $ A_{1g}$   &   $ A_g$   &   $\mathlarger{\sum}_{\textbf{e} \in \{e_x,e_y\} }\bigg(\Delta_{xz}(\textbf{e})+\Delta_{yz}(\textbf{e})\bigg)$    \\ \hline
    $s_{x^2y^2}\ \tau_0$   &   $ A_{1g}$   &   $ A_g$   &   $\mathlarger{\sum}_{\textbf{e} \in \{ e_x \pm e_y\}}\bigg(\Delta_{xz}(\textbf{e})+\Delta_{yz}(\textbf{e})\bigg)$ \\ \hline
    $s_{x^2y^2}\ \tau_z$   &   $ B_{1g}$   &   $ A_g$   &   $\mathlarger{\sum}_{\textbf{e} \in \{ e_x \pm e_y\}}\bigg(\Delta_{xz}(\textbf{e})-\Delta_{yz}(\textbf{e})\bigg)$  \\ \hline
     $d_{x^2-y^2}\ \tau_0$   &    $B_{1g} $   &   $A_g$   &   $\mathlarger{\sum}_{\alpha \in \{xz,yz\} }\bigg(\Delta_{\alpha}(e_x)-\Delta_{\alpha}(e_y)\bigg)$  \\ \hline   
  \end{tabular}
\caption{
Symmetry classification of
spin-singlet intra-orbital pairing channels
by the $D_{4h}$  and $D_{2h}$ point groups. Here,
$\tau_i$ are the Pauli matrices in the $d_{xz},d_{yz}$ orbital basis. A complete list involving these orbitals 
and 
 the
$d_{xy}$ orbital is given in the SM \cite{suppl}}.
\label{tab:1}
\end{table}
\end{center}

{\it Pairing structure in the nematic phase.~}
We next analyze the influence of nematic order on the pairing structure.
The pairing can be classified by the irreducible representations of the point group associated with the lattice symmetry,
which is summarized in Table ~\ref{tab:1} and in the SM \cite{suppl}.
In the tetragonal phase, the corresponding point group is $D_{4h}$. For example, 
the usual $s$-wave 
and 
$d$-wave pairings have an
$A_{1g}$ and a $B_{1g}$
symmetry, respectively. 
In the nematic phase, 
the point group is reduced to $D_{2h}$.
In this case, both the $A_{1g}$ and $B_{1g}$ representations of $D_{4h}$ belong to the $A_g$ representation of the $D_{2h}$ group. As a consequence, the $s$- and $d$-wave pairing channels 
will generically
mix.

We now turn to detailed calculations.
Because the relevant electronic states are dominated by the
$d_{xz}$,$d_{yz}$,and $d_{xy}$ orbitals,
we only consider the nearest-neighbor and next-nearest-neighbor intraorbital exchange interactions for these three orbitals.
As in the previous study of orbital-selective pairing in the tetragonal phase \cite{osp_nica},
we introduce two ratios
$r_L$ and $r_O$. Here,
$r_L=\frac{J_{1}}{J_{2}}$, for each orbital, quantifies the magnetic frustration effect;
$r_O=\frac{J_{2}^{xy}}{J_{2}^{xz/yz}}=\frac{J_{1}^{xy}}{J_{1}^{xz/yz}}$ reflects the orbital-selective effect between the
$xz/yz$ and $xy$ orbitals. 
(The inter-orbital
 pairings
are
negligibly small \cite{osp_nica}.)

\begin{figure}[t!]
\centering\includegraphics[
width=80mm
]{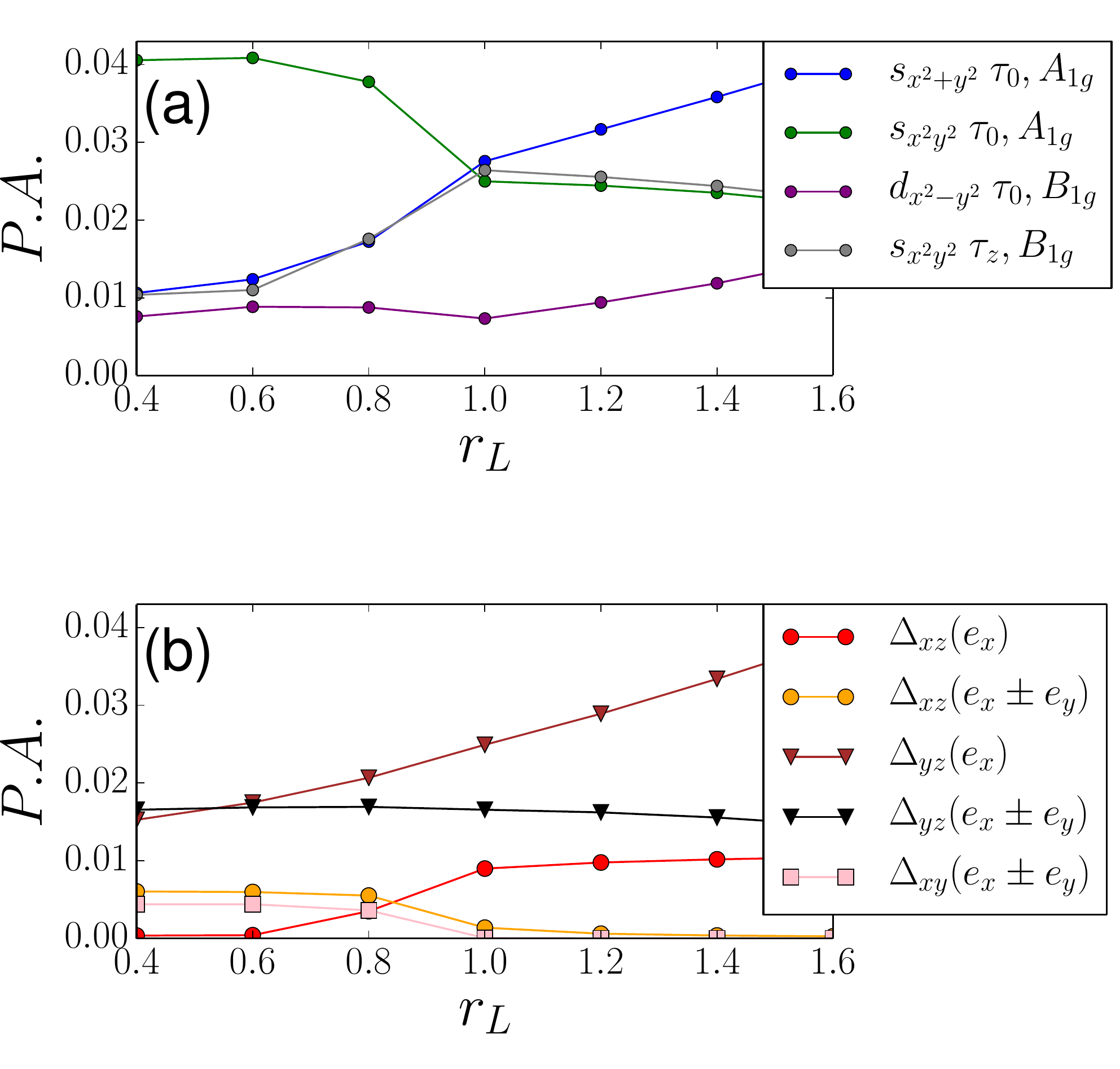}
\caption{(Color online) Top panel: Evolution of the pairing amplitudes
(P.A.), $\Delta$, 
 with magnetic frustration parameter $r_L$ for several channels according to the $D_{4h}$ representations.
 The parameters are $\eta=0.07$, $r_O=0.3$, and $J_{2,xz/yz}=0.3$.
 Bottom panel: Same as the top panel but
shown according to
 the  $D_{2h}$ representations, 
demonstrating a strong orbital-selective pairing with $\Delta_{yz}\gg\Delta_{xz/xy}$.
 }
\label{fig:2}
\end{figure}

In Fig.~\ref{fig:2}, we 
present
the evolution of the pairing amplitudes of several pairing channels with $r_L$.
The top panel shows the pairing channels classified by the $D_{4h}$ group. The dominant pairing always has an $A_{1g}$ symmetry. With increasing $r_L$, it crosses over from the sign-changing $s$ wave (with form factor $\cos k_x \cos k_y$) to an extended $s$-wave (with form factor $\cos k_x + \cos k_y$). It is more transparent to show the pairing amplitudes according to the
irreducible representations of the
$D_{2h}$ group.
 As illustrated in the bottom panel of Fig.~\ref{fig:2},
we find strong orbital-selective pairing with $|\Delta_{yz}|/|\Delta_{xz/xy}|>2$. Such an orbital-selective pairing is quite robust within a wide range of
$r_L$ and $r_O$ values.

The strong orbital selectivity in the superconducting pairing is connected with that of the normal state. To see this, note that from Eq.~\eqref{pairing} we have the ratio of the pairing amplitudes,
\begin{equation}\label{osp}
\frac{|\Delta_{yz}|}{|\Delta_{xz/xy}|}=\frac{Z_{yz}}{Z_{xz/xy}}
\frac{|\tilde{\Delta}_{yz}|}{|\tilde{\Delta}_{xz/xy}|}.
\end{equation}
In other words, the orbital selectivity of the pairing amplitudes 
is magnified by 
$\frac{Z_{yz}}{Z_{xz/xy}}$, the ratio 
of the quasiparticle spectral weights
 in the normal state.

\begin{figure}[t!]
\centering\includegraphics[
width=70mm
]{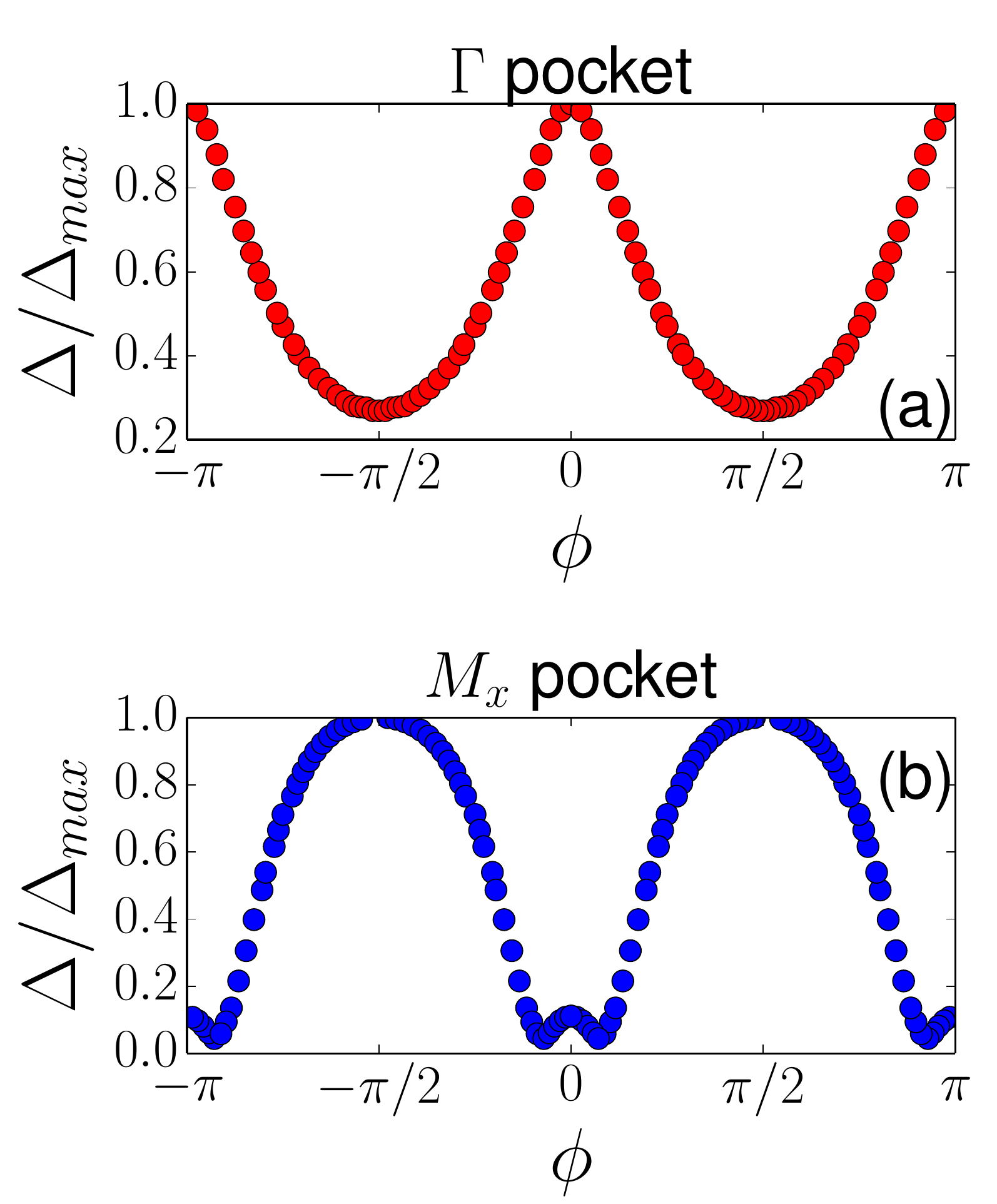}
\caption{Top: Variation of the superconducting gap  on the hole (top panel) and electron (bottom panel) pockets near $\Gamma$ and $M_x$ points of the BZ, respectively. The angle $\phi$ is defined as in Fig.~\ref{fig:1}. Along each pocket,
the gap values are normalized by the corresponding maximum. The calculations are for $r_L=1.2$, $r_O=0.3$, and $\eta=0.07$.}
\label{fig:3}
\end{figure}

{\it Gap anisotropy.~}
We now calculate the superconducting gap on the
normal-state Fermi surface.
In Fig.~\ref{fig:3} we plot the gap variation on the hole (near $\Gamma$) and
 electron (near $M_x$) Fermi pockets. Along each Fermi pocket, the gap values are normalized by its corresponding
  maximal value,
 and the angle $\phi$ is defined in Fig.~\ref{fig:1}.
For the Fermi pocket near $\Gamma$, the gap maximum
appears at $\phi=0/\pi$ and the minimum
is at $\phi=\frac{\pi}{2}$. For the pocket near $M_x$, the maximum is at $\phi=\frac{\pi}{2}$ and
the minimum is close to $\phi=0$.
These positions of the gap maximum/minimum, as well 
as the large gap anisotropy on both Fermi pockets,
 are consistent with the experimental results \cite{stm}. 
More specifically,
i) the ratio of the maximum gap of the hole pocket to that of the electron pocket is of order unity,
about $1.01$ in our calculation. Experimentally, the ratio is comparable to this: It is 
$1.5$ ($1.0$) when the maximal gap on the hole pocket is inferred from the 
STM \cite{stm} (laser-ARPES \cite{Shin-FeSe}) measurements.
ii) The calculated ratio of the gap minimum to gap maximum for the electron pocket ($\sim5\%$)
is comparable to its experimental counterpart (in the range 5\%-30\%) \cite{stm}.
iii) Likewise, the calculated ratio for the hole pocket ($\sim 25 \%$) is comparable to its experimental counterpart 
(4\%-25\%) \cite{stm}.

\begin{figure}[t!]
\centering\includegraphics[
width=70mm
]{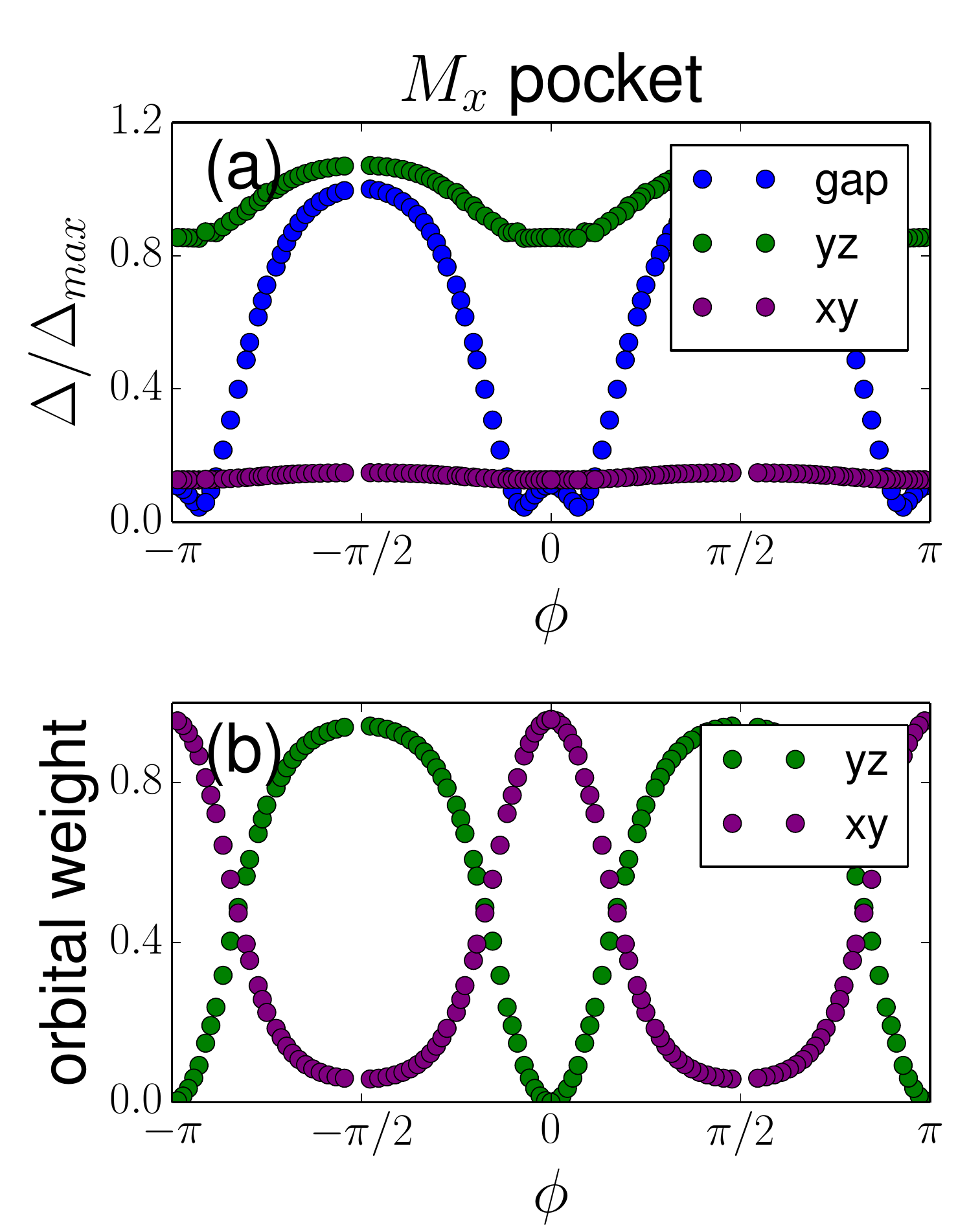}
\caption{(Color online) (Top): Overall and orbital resolved superconducting gaps
along the $M_x$ electron pocket. (Bottom): 
Weight distributions of the $xy$ and $yz$ orbitals along the $M_x$ electron pocket.
}
\label{fig:4}
\end{figure}

Our results are understood as follows.
At any given point of the Fermi surface $\textbf{k}$, the overall gap
$\Delta(\textbf{k})=
\sum \Delta_\alpha (\textbf{k})
W_\alpha(\textbf{k})$.
Here, $W_\alpha$ is the orbital weight, and 
$\Delta_\alpha (\textbf{k})=\sum_{\textbf{e}\in\{e_x,e_y,e_x\pm e_y\}}
J_{\textbf{e}}^{\alpha \alpha} \Delta_\alpha(\textbf{e})\cos(\textbf{k}\cdot \textbf{e}) $
is the orbital-resolved gap.
As an illustration, we show the distributions of 
the orbital-resolved gap
and the corresponding
orbital weight
on the
electron pocket near $M_x$ in Fig.~\ref{fig:4} (and 
for  the hole pocket in the
SM \cite{suppl}).
Along the electron pocket, near $\phi=\frac{\pi}{2}$,
the $yz$ orbital has the largest orbital weight. 
Thus, the gap there is dominated
 by the pairing
in the $yz$ orbital, namely, $\Delta({\phi=\frac{\pi}{2}})\approx \Delta_{yz}$.
Similarly, near $\phi=0$, the $xy$ orbital has the largest orbital weight and then $\Delta({\phi=0})\approx\Delta_{xy}$.
The strong orbital-selectivity in the pairing amplitude  $|\Delta_{yz} (\textbf{e})
|\gg|\Delta_{xy} (\textbf{e})|$
gives rise to
a large gap anisotropy $|\Delta({\phi=\frac{\pi}{2}})|\approx |\Delta_{yz}|\gg |\Delta_{xy}|\approx |\Delta({\phi=\frac{\pi}{2}})|$.
A similar argument applies to the hole pocket,
where $|\Delta({\phi=0})|\approx |\Delta_{yz}|\gg |\Delta_{xz}|\approx |\Delta({\phi=\frac{\pi}{2}})|$, as seen in 
the SM\cite{suppl}.

{\it Discussions.~}
In principle, additional
factors may influence the  gap anisotropy.
 For instance, it has been shown that the magnetic frustration $r_L$ can tune the relative strength of nearest-neighbor and next nearest-neighbor pairings, and gives rise to a moderate level of gap anisotropy along the electron pocket in NaFeAs
 \cite{Yu_PRB2014} .
For FeSe, we have focused on the regime $r_L\sim1$:
The absence of
antiferromagnetic 
order in the nematic state
 suggests a strong magnetic frustration with $r_L\sim1$,
where the nearest-neighbor and next nearest-neighbor pairings are quasidegenerate.

In the calculations we have carried out, the nematicity has multiple effects on the pairing structure.
First, it enhances the orbital selectivity in the spectral weight of the coherent itinerant electrons, 
leading to strong orbital-selective pairing amplitudes,
as shown in Eq.~\eqref{osp}. Second, the orbital weights are largely redistributed along the distorted Fermi surface
as a combined effect of the additional anisotropy and orbital-dependent band-structure renormalization in the nematic phase.
On each Fermi pocket, the dominant orbital character has a large variation.
Third, the nematicity induces additional magnetic anisotropy, which
enhances the pairing in the $e_x$ direction but 
reduces the pairing in the $e_y$ direction.
While this last effect also contributes to the gap anisotropy, it is not the dominant source in our case.
In other words, the gap anisotropy primarily originates from the first two effects,
which 
dictate 
the orbital-selective nature of the pairing amplitudes.

The orbital-selective pairing concerns superconductivity driven by short-range spin-exchange interactions
between the electrons associated with the multiple $3d$ orbitals.
For FeSe, direct evidence exists that the local Coulomb (Hubbard and Hund's) interactions are strong \cite{Watson2017,Evtushinksy},
and the orbitals thus represent a natural basis to consider superconducting pairing.

We now discuss 
the broader implications of the orbital selective pairing. 
There is accumulating evidence that superconductivity in the FeSCs is mainly driven by magnetic correlations.
Yet, the precise role of the nematicity on the superconductivity remains an open question.
Our study raises the possibility that the main influence of the nematicity on the magnetically driven superconductivity 
is through
its influence on the orbital selectivity.

Finally, the correlation effects provide 
intuition on how to control low-energy physics 
by tuning local degrees of freedom. For instance,  the multi-orbital nature affords a handle for engineering the low-energy 
electronic states and raising $T_c$. Even when the superconductivity is primarily driven by magnetic correlations, 
tuning the orbital levels and orbital-dependent couplings may optimize superconductivity. This notion is consistent with 
experiments on single-layer 
FeSe
 \cite{XShi},
which indicate a further increased $T_c$ 
by varying 
the weight of particular $3d$ orbitals
near the Fermi energy.

{\it Conclusions.~}
We have studied the superconductivity
in the nematic phase of FeSe
 through a
multiorbital model 
using a $U(1)$ slave-spin approach. 
The enhanced orbital selectivity
in the normal state by the nematic order 
is shown to yield 
a strong orbital-selective superconducting pairing.
The latter produces sizable gap anisotropy on both the hole and electron pockets,
which naturally explains 
the recent experimental observations.
The orbital-selective pairing
  raises the prospect of harnessing the orbital degrees of freedom to
realize still higher $T_c$,  even when superconductivity is magnetically driven,
and 
provides insights into the interplay
between electronic orders and superconductivity. As such, 
our results 
shed light 
not only on the physics of the iron-based 
compounds but also on
the unconventional
superconductivity in a variety of  other strongly correlated systems.

\acknowledgements
We thank E. Abrahams, S. V.
Borisenko, J. C. S. Davis, W. X. Ding, and X.-J. Zhou 
for useful discussions. The work has in part been supported 
by the U.S. Department of Energy, Office of Science, Basic 
Energy Sciences, under Award No. DE-SC0018197 and
 the Robert A. Welch Foundation Grant No. C-1411 (H.H.
 and Q.S.), by the National Science Foundation of China
 Grant No. 11674392, Ministry of Science and Technology
 of China, National Program on Key Research Project Grant
 No. 2016YFA0300504 and the Research Funds of Remnin
 University of China Grant No. 18XNLG24 (R.Y.), by ASU
  Startup Grant (E.M.N.), by the U.S. DOE Office of Basic
 Energy Sciences E3B5 (J.-X.Z.). It was also in part supported
  by the Center for Integrated Nanotechnologies, a U.S. 
DOE BES user facility. Q.S. acknowledges the support of 
ICAM and a QuantEmX grant from the Gordon and Betty 
Moore Foundation through Grant No. GBMF5305 (Q.S.), the 
hospitality of University of California at Berkeley and of the 
Aspen Center for Physics (NSF Grant No. PHY-1607611), 
and the hospitality and the support by a Ulam Scholarship 
from the Center for Nonlinear Studies at Los Alamos National 
Laboratory.

\clearpage
\setcounter{figure}{0}
\makeatletter
\renewcommand{\thefigure}{S\@arabic\c@figure}
\onecolumngrid
\section{Supplemental Material}

\subsection{Details on the Model and Self-consistent Calculations of the Pairing Amplitudes}
We consider a five-orbital Hubbard model for FeSe. The Hamiltonian reads as
\begin{eqnarray}\label{HamHubbard}
H&=&H_{t}+H_{\rm{int}},\nonumber\\
H_{t}&=&\sum_{ij,\alpha\beta}t_{ij}^{\alpha\beta} c_{i,\alpha,\sigma}^\dag c_{j,\beta,\sigma},\nonumber\\
 H_{\rm{int}} &=& \frac{U}{2} \sum_{i,\alpha,\sigma}n_{i\alpha\sigma}n_{i\alpha\bar{\sigma}} 
 +\sum_{i,\alpha<\beta,\sigma} \left\{ U^\prime n_{i\alpha\sigma} n_{i\beta\bar{\sigma}}\right.
 + (U^\prime-J_{\rm{H}}) n_{i\alpha\sigma} n_{i\beta\sigma}\nonumber\\
&&\left.-J_{\rm{H}}(c^\dagger_{i\alpha\sigma}c_{i\alpha\bar{\sigma}} c^\dagger_{i\beta\bar{\sigma}}c_{i\beta\sigma}
 +c^\dagger_{i\alpha\sigma}c^\dagger_{i\alpha\bar{\sigma}}
 c_{i\beta\sigma}c_{i\beta\bar{\sigma}}) \right\},
\end{eqnarray}
where $c_{i,\alpha,\sigma}^\dag$ creates an electron in orbital $\alpha(\in{xz,yz,x^2-y^2,xy,z^2})$, spin $\sigma$
and at site $i$ of an Fe-square lattice,
and
$H_{\rm{int}}$ describes the on-site interactions,
which include the intra- and inter-orbital Coulomb repulsions $U$ and $U^\prime$,
and the Hund's coupling $J_{\rm{H}}$. After integrating out the incoherent part of the electron spectrum via the $U(1)$ slave-spin approach, we obtain an effective multi-orbital $t$-$J$ model. In terms of the $f$-fermions this effective model
takes the following form:
\begin{eqnarray}
H_{eff}&=&\sum_{ij,\alpha\beta}\widetilde{t}_{ij}^{\alpha\beta} f_{i,\alpha,\sigma}^\dag f_{j,\beta,\sigma} 
-\sum_{ij,\alpha\beta} J_{ij}^{\alpha\beta}f_{j,\beta,\downarrow}^\dag f_{i,\alpha,\uparrow}^\dag f_{i,\alpha,\downarrow}f_{j,\beta,\downarrow}
\end{eqnarray}
with the normalized hopping term $\widetilde{t}_{ij}^{\alpha\beta}=\sqrt{Z_{\alpha}Z_{\beta}}t_{ij}^{\alpha\beta}-\lambda_{\alpha}\delta_{\alpha\beta}\delta_{ij}$. Here we only consider the intraorbital exchange interaction with $J^{\alpha\beta}=J^{\alpha}\delta_{\alpha\beta}$.

We
consider
 the pairing amplitudes of the $f$-fermions,
\begin{eqnarray}
\widetilde{\Delta}_{\textbf{e},\alpha}=\frac{1}{2N}\sum_i\bigg<f_{i,\alpha,\uparrow}f_{i+\textbf{e},\alpha,\downarrow}-f_{i,\alpha,\downarrow}f_{i+\textbf{e},\alpha,\uparrow}\bigg>
\end{eqnarray}
where $\textbf{e}\in\{e_x,e_y,e_{x+y},e_{x-y}\}$. From a
Hubbard-Stratonovich decoupling,
the $H_{eff}$ is reduced into fermion bilinears and
 $\widetilde{\Delta}_{\textbf{e},\alpha}$ can then be solved.
The pairing amplitude of the physical electrons is obtained as follows:
\begin{eqnarray}
{\Delta}_{\textbf{e},\alpha}=\frac{1}{2N}\sum_i\bigg<c_{i,\alpha,\uparrow}c_{i+\textbf{e},\alpha,\downarrow}-c_{i,\alpha,\downarrow}c_{i+\textbf{e},\alpha,\uparrow}\bigg>=Z_{\alpha}\widetilde{\Delta}_{\textbf{e},\alpha}
\end{eqnarray}

\begin{figure}[h!]
\centering\includegraphics[
width=140mm
]{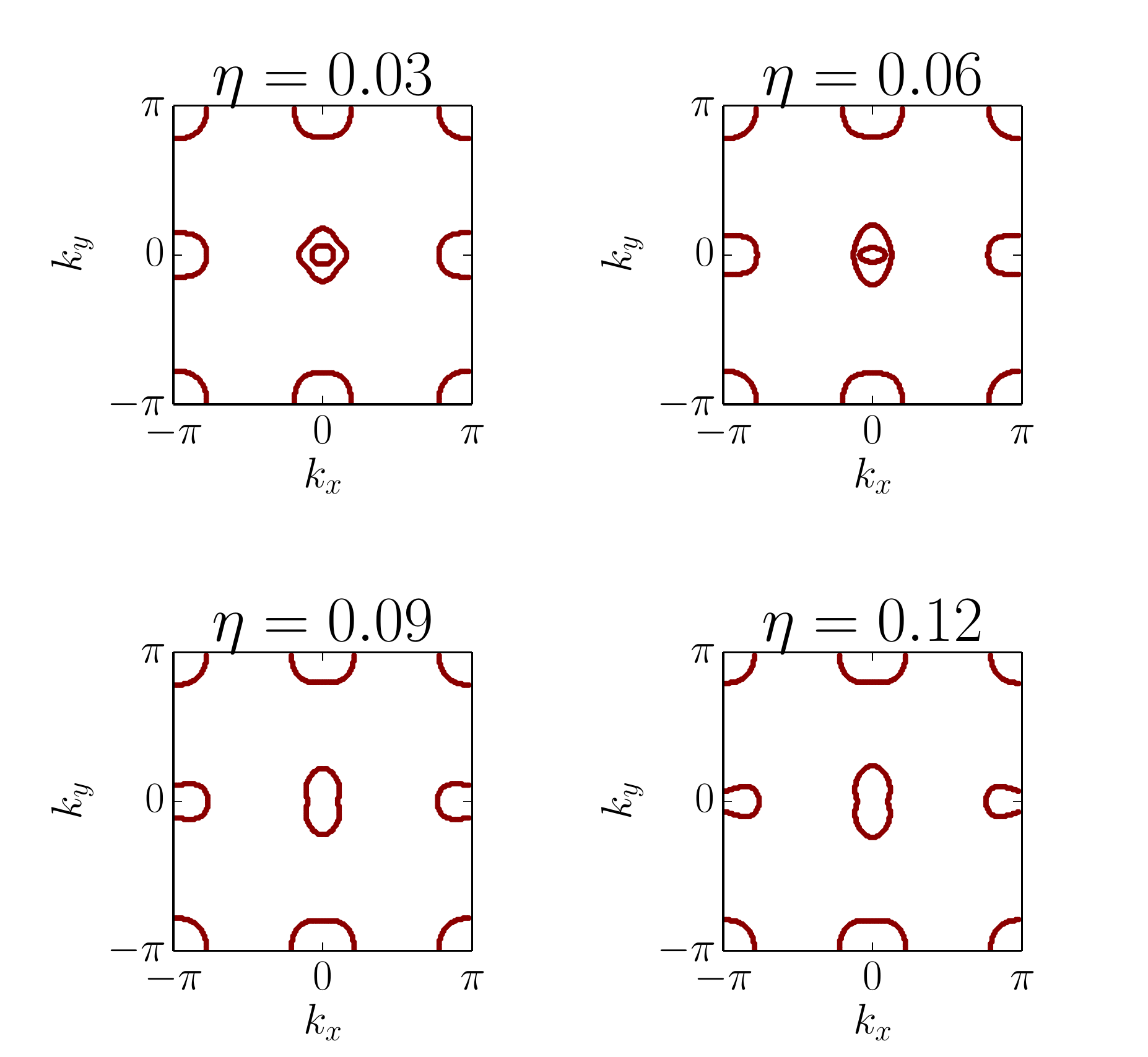}
\caption{(Color online) Evolution of the Fermi surface with the anisotropy parameter $\eta$ in the nematic phase.}
\label{Sfig:1}
\end{figure}

\begin{figure}[h!]
\centering\includegraphics[
width=140mm
]{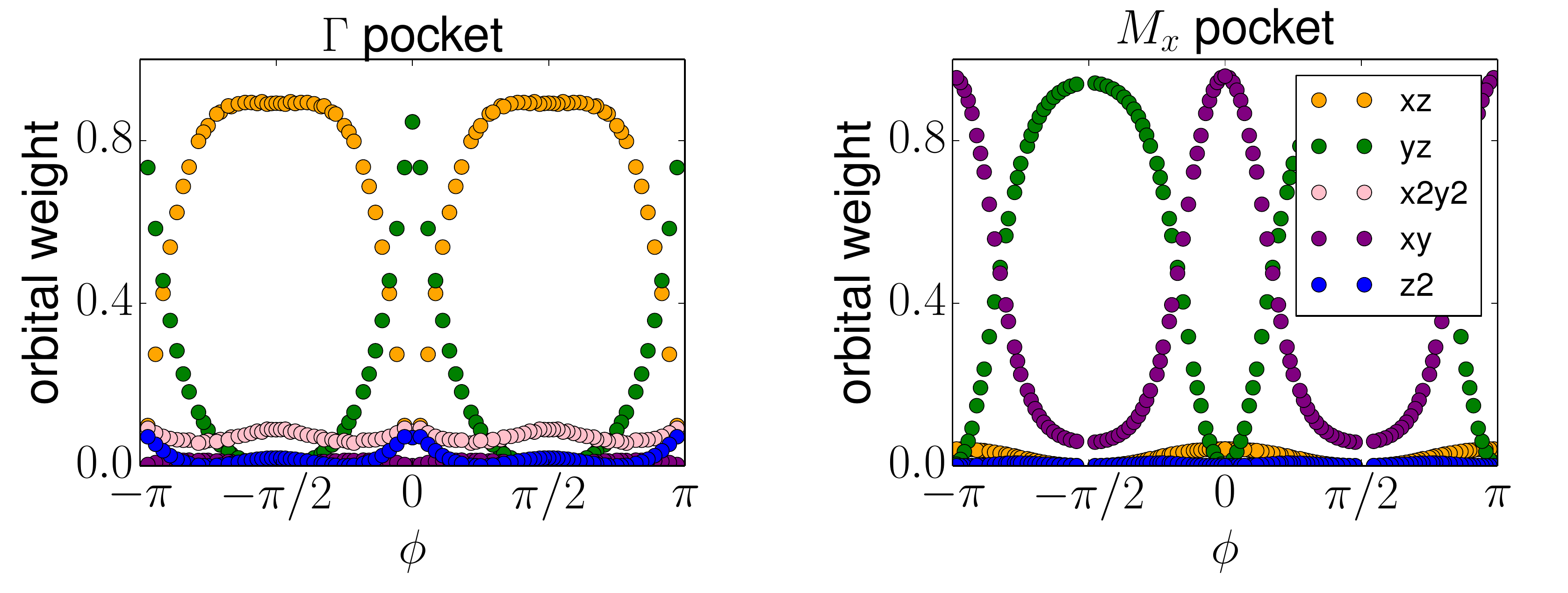}
\caption{(Color online) Orbital weights along the outer hole pocket near $\Gamma$ (left panel) and those for the 
electron pocket near M$_x$ (right panel).}
\label{Sfig:2}
\end{figure}

\subsection{Details on the tight-binding parameters}
We present the tight-binding parameters in Table S1. To obtain the tight-binding parameters, we perform local density approximation (LDA) calculations for bulk FeSe
with a tetragonal structure, and we fit the LDA band structure to the tight-binding Hamiltonian.
The form of the five-orbital tight-binding Hamiltonian given in Ref.~[\onlinecite{SM_Graser_2009}]
is used. 

\begin{minipage}{\linewidth}
\begin{center}
\begin{tabular}{cccccccc}
  \hline
  \hline
    & $\alpha=1$ & $\alpha=2$ & $\alpha=3$ & $\alpha=4$ & $\alpha=5$ &   &   \\ \hline
  $\epsilon_\alpha$ & -0.00733 & -0.00733 & -0.52154 & 0.10974 & -0.5694 &  & \\ \hline\hline
  $t^{\alpha\alpha}_\mu$ & $\mu=x$ & $\mu=y$ & $\mu=xy$ & $\mu=xx$ & $\mu=xxy$ & $\mu=xyy$ & $\mu=xxyy$ \\ \hline
  $\alpha=1$ & -0.0111 & -0.49155 & -0.23486 & -0.0119 & -0.04025 & -0.03917 & -0.03808\\ \hline
  $\alpha=3$ & -0.38485 &  & -0.08015 & -0.00646 &  &  &  \\ \hline
  $\alpha=4$ & -0.16872 &  & -0.10728 & -0.00626 & -0.04592 &  & -0.02079 \\ \hline
  $\alpha=5$ & -0.03681 &  &  & -0.00159 & -0.01585 &  & -0.02739 \\ \hline\hline
  $t^{\alpha\beta}_\mu$ & $\mu=x$ & $\mu=xy$ & $\mu=xxy$ & $\mu=xxyy$ &  &  &  \\ \hline
  $\alpha\beta=12$ &  & -0.12701 & -0.00655 & -0.05869 &  &  &  \\ \hline
  $\alpha\beta=13$ & -0.36123 & -0.07201 & -0.0134 &  &  &  &  \\ \hline
  $\alpha\beta=14$ & -0.20068 & -0.03548 & -0.00705 &  &  &  &  \\ \hline
  $\alpha\beta=15$ & -0.08057 & -0.14823 &  & -0.01218 &  &  &  \\ \hline
  $\alpha\beta=34$ &  &  & -0.0217 &  &  &  &  \\ \hline
  $\alpha\beta=35$ & -0.29868 &  & -0.01332 &  &  &  &  \\ \hline
  $\alpha\beta=45$ &  & -0.13208 &  & -0.05213 &  &  &  \\ \hline
  \hline
\end{tabular}
\par
\end{center}
\bigskip
\noindent
{\bf Supplemental Table S1}.
 Tight-binding parameters of the five-orbital model for
 bulk FeSe with the tetragonal structure.
Here we
use the same notation as in Ref.~
[\onlinecite{SM_Graser_2009}].
The orbital index $\alpha=$1,2,3,4,5 correspond to $d_{xz}$, $d_{yz}$,
$d_{x^2-y^2}$, $d_{xy}$, and $d_{3z^2-r^2}$ orbitals, respectively.
The listed parameters are in eV.
\end{minipage}

\subsection{Evolution of the Fermi surface}

To study the effect of nematic order on the Fermi surface, we set the interaction strength to zero and calculate the Fermi surface and orbital weight distribution of the tight-binding model with various nematic order parameters. In Fig.~\ref{Sfig:1},
we show the evolution of the Fermi surface with the anisotropy parameter $\eta$. 
The  $\Gamma$ pocket enlarges
in the $y$ direction and both $\Gamma$ and M$_x$ pockets have a peanut shape for sufficiently large $\eta$. 
In Fig.~\ref{Sfig:2},
we plot the orbital weight distribution at $\eta=0.07$.
The $\Gamma$ pocket is dominated by the $d_{xz}$ and $d_{yz}$ orbitals
and the M$_x$ pocket
by the $d_{yz}$ and $d_{xy}$ orbitals.

\subsection{Classification of the Pairing Channels}

The pairing channels can be classified by
using the irreducible representations of the lattice point group of the system.
For the tetragonal symmetry, the corresponding lattice point group is $D_{4h}$.
In the presence of the nematic order,
the $C_4$ rotational symmetry is broken, and the
point group
is reduced
to $D_{2h}$. In the main text, we consider the intraorbital spin-singlet pairing channels of the system.
A full list of these pairing channels involving the $d_{xz}$, $d_{yz}$, and $d_{xy}$ orbitals and 
their corresponding symmetry classification are 
given in Table S2.

\begin{minipage}{\linewidth}
\begin{center}
  \begin{tabular}{ | l | l | l | l | }
    \hline
      pairing channel in momentum space &  $D_{4h}$  &  $D_{2h}$   &   pairing channel in real space \\ \hline
    $s_{x^2+y^2}\tau_0$   &   $ A_{1g}$   &   $ A_g$   &   $\Delta_{xz}(e_x)+\Delta_{xz}(e_y)+\Delta_{yz}(e_x)+\Delta_{yz}(e_y)$    \\ \hline
    $s_{x^2y^2}\tau_0$   &   $ A_{1g}$   &   $ A_g$   &   $\Delta_{xz}(e_x+e_y)+\Delta_{xz}(e_x-e_y)+\Delta_{yz}(e_x+e_y)+\Delta_{yz}(e_x-e_y)$  \\ \hline
    $d_{x^2-y^2}\tau_z$   &   $ A_{1g}$   &   $ A_g$   &    $\Delta_{xz}(e_x)-\Delta_{xz}(e_y)-\Delta_{yz}(e_x)+\Delta_{yz}(e_y)$\\ \hline
    $d_{x^2-y^2}\tau_0$   &    $B_{1g} $   &   $A_g$   &   $\Delta_{xz}(e_x)-\Delta_{xz}(e_y)+\Delta_{yz}(e_x)-\Delta_{yz}(e_y)$  \\ \hline
     $s_{x^2+y^2}\tau_z$   &   $B_{1g} $   &   $A_g$   &    $\Delta_{xz}(e_x)+\Delta_{xz}(e_y)-\Delta_{yz}(e_x)-\Delta_{yz}(e_y)$ \\ \hline
     $s_{x^2y^2}\tau_z$   &   $B_{1g}$   &   $A_g $   &    $\Delta_{xz}(e_x+e_y)+\Delta_{xz}(e_x-e_y)-\Delta_{yz}(e_x+e_y)-\Delta_{yz}(e_x-e_y)$  \\ \hline
     $d_{xy}\tau_z$   &   $A_{2g}$   &   $B_{1g}$   &  $\Delta_{xz}(e_x+e_y)-\Delta_{xz}(e_x-e_y)-\Delta_{yz}(e_x+e_y)+\Delta_{yz}(e_x-e_y)$   \\ \hline
     $d_{xy}\tau_0$   &   $ B_{2g} $   &   $ B_{1g}$   &    $\Delta_{xz}(e_x+e_y)-\Delta_{xz}(e_x-e_y)+\Delta_{yz}(e_x+e_y)-\Delta_{yz}(e_x-e_y)$   \\ \hline
     $s_{x^2y^2} \mathds{1}_{xy}$   &   $ A_{1g}$   &   $ A_g$   &   $\Delta_{xy}(e_x+e_y)+\Delta_{xy}(e_x-e_y)$    \\ \hline
    $s_{x^2+y^2} \mathds{1}_{xy}$   &   $ A_{1g}$   &   $ A_g$   &    $\Delta_{xy}(e_x)+\Delta_{xy}(e_y)$  \\ \hline
    $d_{x^2-y^2} \mathds{1}_{xy}$   &   $ B_{1g}$   &   $ A_g$   &   $\Delta_{xy}(e_x)-\Delta_{xy}(e_y)$  \\ \hline
    $d_{xy} \mathds{1}_{xy}$   &    $B_{2g} $   &   $B_{1g}$   &   $\Delta_{xy}(e_x+e_y)-\Delta_{xy}(e_x-e_y)$  \\ \hline

  \end{tabular}
\par
\end{center}
\bigskip
\noindent
{\bf Table S2}.
Symmetry classification of
the spin-singlet intra-orbital
pairing channels
involving the \text{$d_{xz},d_{yz},d_{xy}$} orbitals. Here \text{$\tau_i$} are the Pauli matrices of the isospin operator in the \text{$d_{xz/yz}$} orbital basis.
\end{minipage}

\subsection{Phase Diagram and the Pairing Amplitude}

\begin{figure}[h!]
\centering\includegraphics[
width=140mm
]{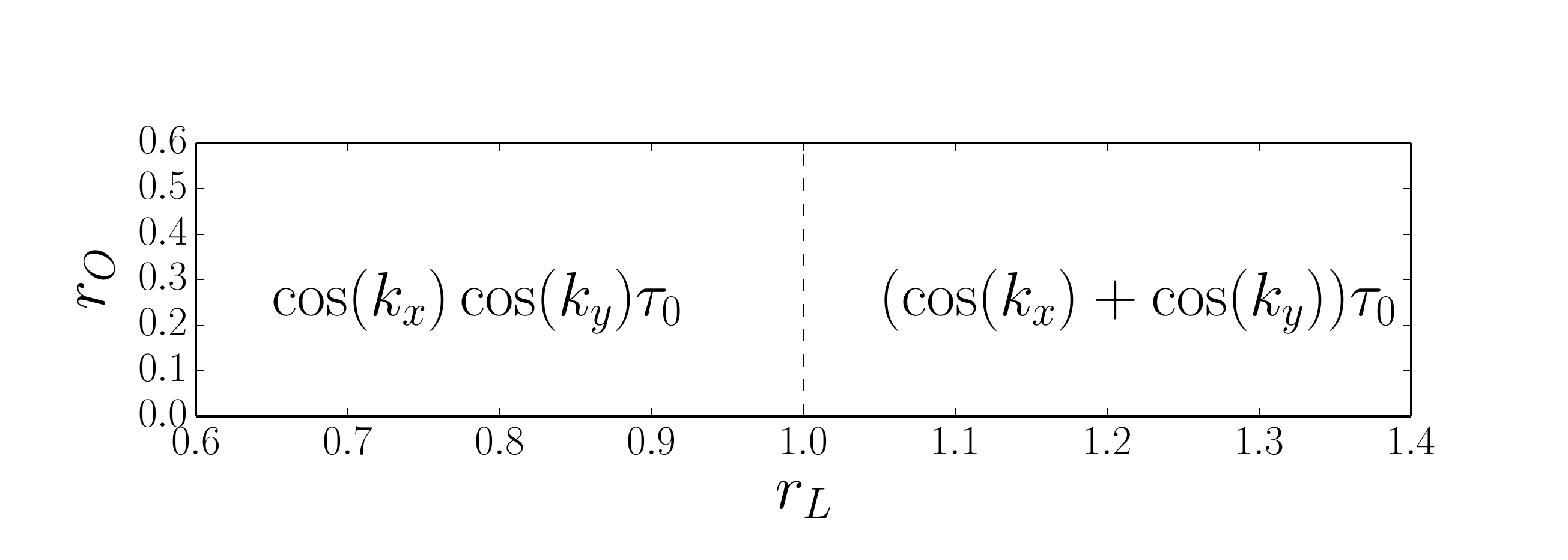}
\caption{(Color online) Phase diagram 
showing the different regimes with 
different leading 
pairing channels in the $r_L$-$r_O$ plane.}
\label{Sfig:3}
\end{figure}

\begin{figure}[h!]
\centering\includegraphics[
width=140mm
]{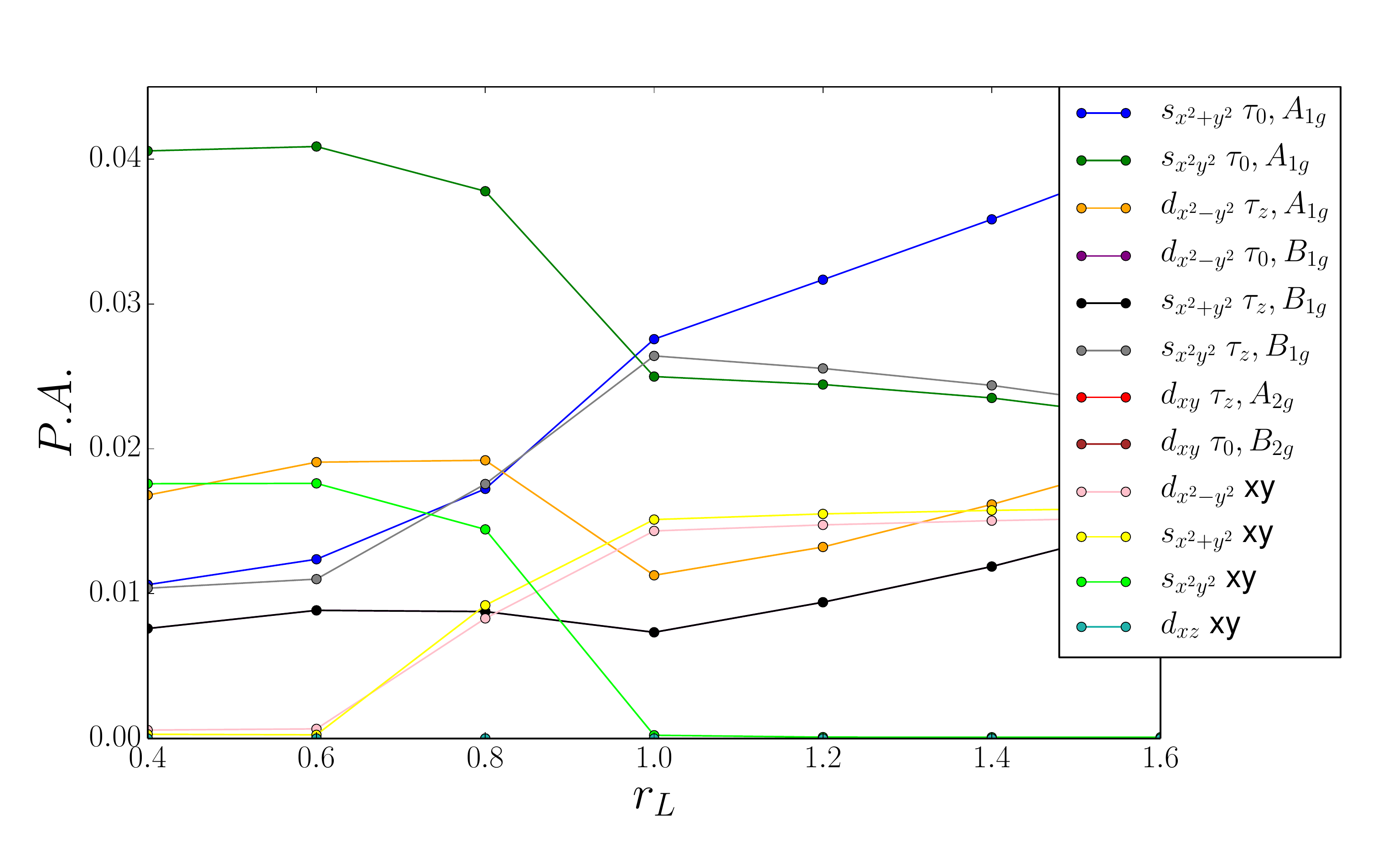}
\caption{(Color online) Pairing amplitudes in the different channels according to the irreducible representations of the $D_{4h}$ group.}
\label{Sfig:4}
\end{figure}

\begin{figure}[h!]
\centering\includegraphics[
width=140mm
]{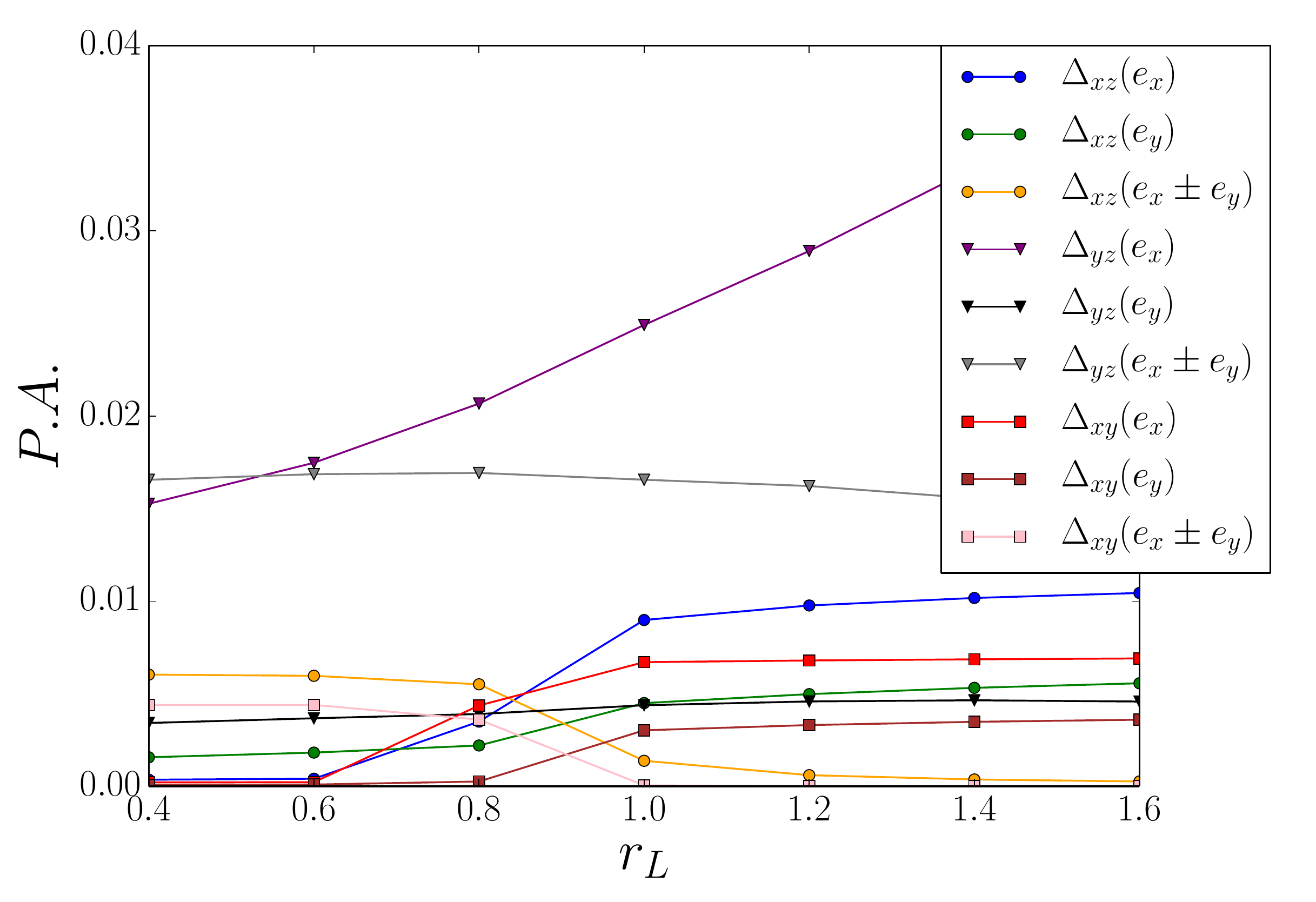}
\caption{(Color online) Pairing amplitudes of the different orbitals
expressed in the 
irreducible representations of the $D_{2h}$ group.
}
\label{Sfig:5}
\end{figure}

To study the 
evolution of 
pairing symmetry
 in the nematic phase, we fix $\eta=0.07$
  and solve for the pairing amplitudes 
at different $r_L$ and $r_O$ values.
Fig.~\ref{Sfig:3} is the resulting phase diagram where each 
regime  is characterized by the
leading pairing channel. Fig.~\ref{Sfig:4} and Fig.~\ref{Sfig:5} show the evolution of pairing amplitude with $r_L$ at $r_O=0.3$.
As is seen,
the pairing amplitude in the $d_{yz}$ orbital is always larger than 
those of
the $d_{xz}$ and $d_{xy}$ orbitals. In addition, the pairing in $e_x$ direction 
is dominant when $r_L$ is small and the pairing in $e_{x\pm y}$ direction 
is dominant when  $r_L$ is large.

\subsection{
Gap anisotropy along the 
$\Gamma$ pocket
}

\begin{figure}[h!]
\centering\includegraphics[
width=140mm
]{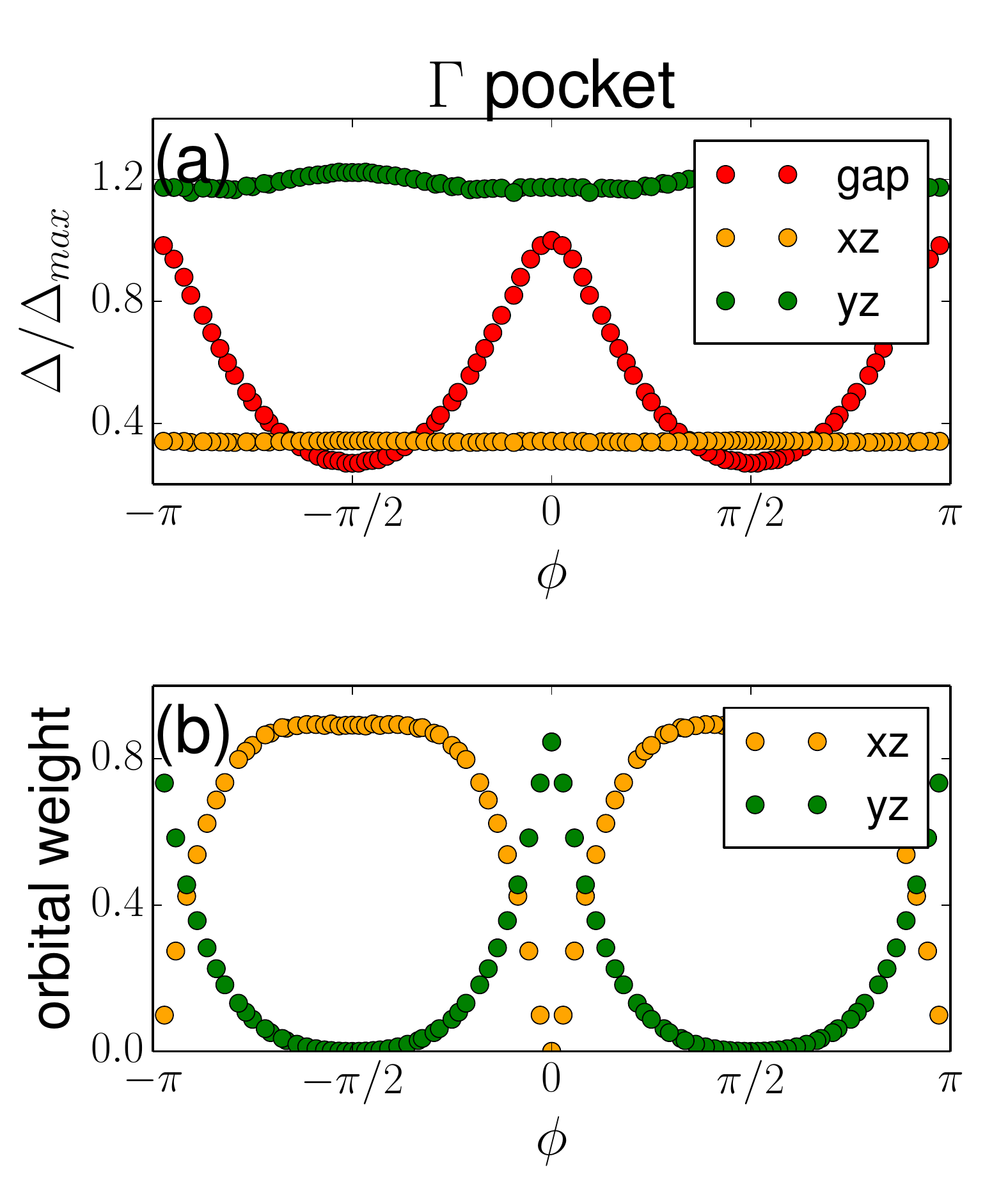}
\caption{(Color online) Gap anisotropy and orbital weight distribution of the 
$\Gamma$ pocket}
\label{Sfig:6}
\end{figure}

In Fig.~\ref{Sfig:6} (top panel), we show the gap anisotropy and the paring strength of the $d_{xz}$ and $d_{yz}$ orbitals along the
$\Gamma$ hole pocket. In Fig.~\ref{Sfig:6} (bottom panel), we plot the weight distributions of the $d_{xz}$ and $d_{yz}$ orbitals along the
same pocket.
At $\phi=0$, the $yz$ orbital has the largest orbital weight, and $\Delta({\phi=0})\sim {\Delta}_{yz}$.
At $\phi=\frac{\pi}{2}$, the $xz$ orbital has the largest orbital weight; correspondingly,
$\Delta ({\phi=\frac{\pi}{2}}) \sim \Delta_{xz}$.
For strong orbital selective pairing,
$\Delta_{yz} \gg \Delta_{xz}$.
As a result, the gap will become very anisotropic with $\Delta ({\phi=0} ) \gg \Delta ({\phi=\frac{\pi}{2}})$.

\end{document}